%% file: driver_jfm.tex
\documentclass{jfm}

\usepackage{graphicx}
\usepackage{newtxtext}
\usepackage{newtxmath}
\usepackage{natbib}
\usepackage{hyperref}
\usepackage{subcaption}
\usepackage{bm}
\usepackage{siunitx}

\hypersetup{
    colorlinks = true,
    urlcolor   = blue,
    citecolor  = black,
}

\newcommand{\eone}{$\boldsymbol{e}_1 \,$}
\newcommand{\etwo}{$\boldsymbol{e}_2 \,$}
\newcommand{\etri}{$\boldsymbol{e}_3 \,$}
\newcommand{\lone}{$\lambda_1 \,$}
\newcommand{\ltwo}{$\lambda_2 \,$}
\newcommand{\ltri}{$\lambda_3 \,$}
\newcommand{\w}{$\boldsymbol{\omega} \;$}
\newcommand{\s}{$\mathsfbi{S} \;$}
\newcommand{\A}{$\mathsfbi{A} \;$}
\newcommand{\rev}[1]{\textcolor{black}{#1}}

\newcommand{\drp}{$\mathcal{D}_{\mathbf{\vee}}$}
\newcommand{\drn}{$\mathcal{D}_{\mathbf{\wedge}}$}
\newcommand{\mdrp}{$\overline{\mathcal{D}_{\mathbf{\vee}}}$}
\newcommand{\mdrn}{$\overline{\mathcal{D}_{\mathbf{\wedge}}}$}
\newcommand{\RomanNumeralCaps}[1]

\title{Experimental measurement of the vorticity-strain alignment around extreme energy transfer events}

\author{Benjamin Musci\aff{1}\corresp{\email{benjamin.musci@cea.fr}}, 
    B\'ereng\`ere Dubrulle\aff{1}
   , 
   Jean LeBris\aff{1} ,
   Damien Geneste\aff{1} ,
    Pierre Braganca\aff{2} ,
   Jean-Marc Foucaut\aff{2} ,
   Christophe Cuvier\aff{2} ,
 \and  Adam Cheminet\aff{1}}

\affiliation{\aff{1}Université Paris-Saclay, CEA, CNRS, SPEC, CEA Paris-Saclay, F-91191 Gif sur Yvette, France
\aff{2}Univ. Lille, CNRS, ONERA, Arts et Metiers Institute of Technology, Centrale Lille, UMR 9014 - Laboratoire de Mécanique des Fluides de Lille (LMFL) - Kampé de Fériet, Lille, F-59000, France 
}

\begin{document}
\maketitle

\begin{abstract}
This work experimentally explores the alignment of the vorticity vector and the strain-rate tensor eigenvectors at locations of extreme upscale and downscale energy transfer. We show that the turbulent von Karman flow displays vorticity-strain alignment behavior across a large range of Reynolds numbers, which are very similar to previous studies on homogeneous, isotropic turbulence. We observe that this behavior is amplified for the largest downscale energy transfer events, which tend to be associated with sheet-like geometries. These events are also shown to have characteristics previously associated with high flow field non-linearity and singularities. In contrast the largest upscale energy transfer events display much different structures which showcase a strong preference for vortex-compression. We then show further evidence to the argument that strain-self-amplification is the most salient feature in characterizing cascade direction. Finally, we identify possible invariant behavior for the largest energy transfer events, even at scales near the Kolmogorov scale. 

\end{abstract}

\begin{keywords}
Authors should not enter keywords on the manuscript, as these must be chosen by the author during the online submission process and will then be added during the typesetting process (see \href{https://www.cambridge.org/core/journals/journal-of-fluid-mechanics/information/list-of-keywords}{Keyword PDF} for the full list).  Other classifications will be added at the same time.
\end{keywords}

{\bf MSC Codes }  {\it(Optional)} Please enter your MSC Codes here

\input{sections/intro}
\input{sections/exp_facility}

\input{sections/results}
\input{sections/discussion}



\backsection[Acknowledgements]{The authors thank  C\'{e}cile Wiertel-Gasquet for her countless efforts in the coordination and execution of the experiments, as well as her patience with the first author. They also thank Vincent Padilla for his practical wizardry and quick problem solving. Thanks are also due to Alain Pumir for useful discussions on the interpretation of these results. Erin Wilson also deserve thanks for her assistance with the schematics in this work.  Finally, the authors are immensely thankful to Mauro Abella for his freight elevator rescue and enjoyable lunchtime discussions.}


\backsection[Funding]{This work received support from the Ecole Polytechnique, CNRS, from ANR TILT (grant No. ANR-20-CE30-0035) and from ANR BANG (grant No. ANR-22-CE30-0025). This project has also received funding from the European Union's Horizon 2020 research and innovation program under the Marie Sklodowska-Cure grant agreement No. 945298.}

\backsection[Declaration of interests]{The authors report no conflict of interest.}

\backsection[Author ORCIDs]{B. Musci, https://orcid.org/0000-0002-2693-0543; A. Cheminet, https://orcid.org/0000-0002-0702-9194; C. Cuvier, https://orcid.org/0000-0001-6108-6942; P. Braganca, https://orcid.org/0009-0005-5290-1913; D. Geneste, https://orcid.org/0009-0007-7341-2590; JM. Foucaut, https://orcid.org/0000-0003-0800-8608; B. Dubrulle, https://orcid.org/0000-0002-3644-723X }


\bibliographystyle{jfm}
\bibliography{condd_bib}

\end{document}

%% file: sections/intro.tex
\section{\label{sec:intro}Introduction}

In turbulence there is little question that an energy cascade exists, whereby energy injected at the largest scales of a flow is eventually dissipated to heat by viscosity. However, there is still much to debate as to how this cascade works - the relevant mechanisms and flow structures, the  validity of the so called ``Richardson" cascade, and the degree of time and space locality in the cascade - are all open questions \citep{Tsinober2009, Vela-Martin2021,Johnson2024}. 

\begin{figure}
\centerline{\includegraphics[width=0.9\linewidth]{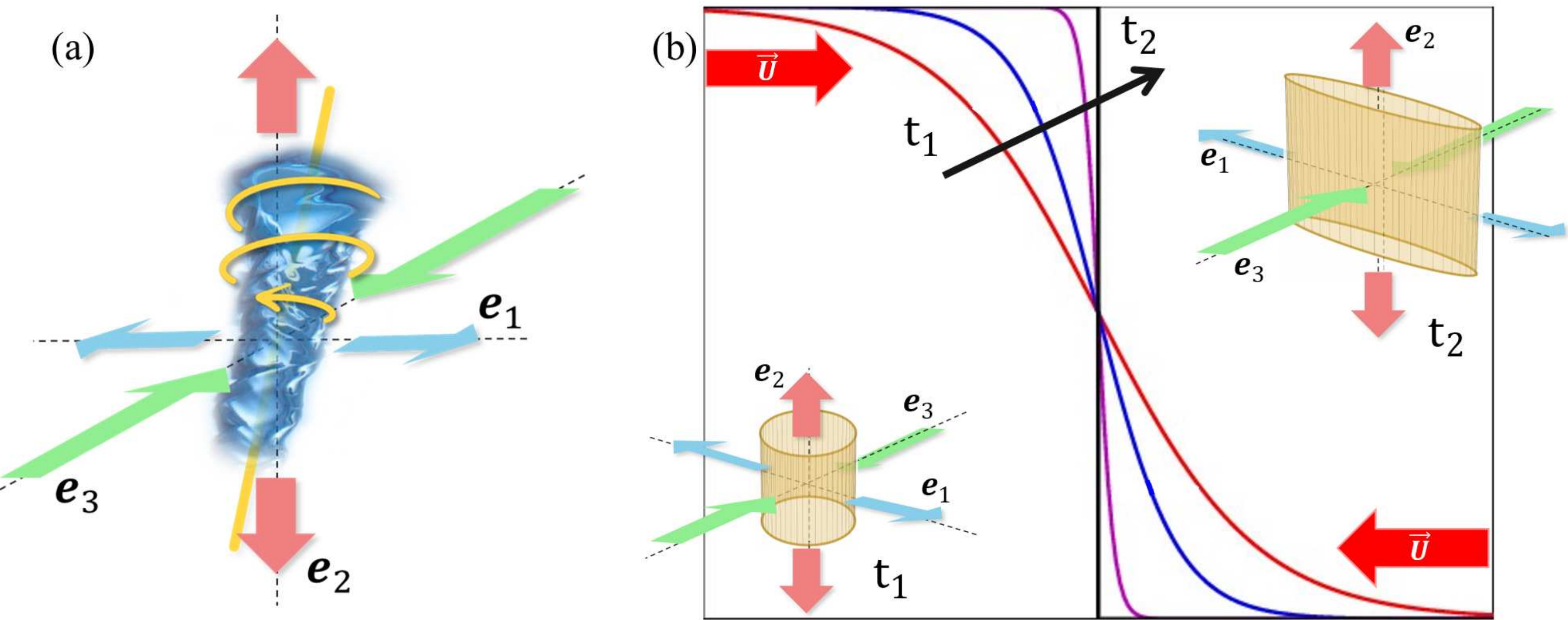}}
\caption{\label{fig:alignment} (a) Schematic of typical  \w- \s alignment. \w is indicated by the yellow line and swirl. The baseline/universal preference for alignment between \w- \etwo is shown. To maximize vortex-stretching, \w (yellow line) would be aligned more with \eone, which has been found generally to not be the case in turbulence. (b) Schematic illustrating strain-rate self-amplification (SSA) in a 1D and 3D sense. In 1D, an initial compressive strain (red line) causes the opposing velocities to steepen the negative gradient of the curves as time progresses, which in turn increases the compressive strain, until a singularity/shock forms (black line) as in the Burgers 1D equation. The same idea is illustrated in the 3D, where an initially cylindrical fluid parcel (or vortex) is compressed along one axis, and stretched in two others, in turn increasing the compressive strain.}
\end{figure}

Perhaps motivated by Richardson's ``whirl" centric poetic verse, a large portion of previous works focus on probing the cascade primarily through the characterization of the vorticity vector, \w\!,  and secondly via the strain-rate tensor, \s\!, which is simply the symmetric portion of the velocity-gradient tensor, \A = $  \nabla \bm{u} = \partial u_i / \partial x_j $. \,
Much of the literature has focused on the alignment and interaction of $\omega_i = \epsilon_{ijk} \partial{u_k} / \partial{x_j}$ and the eigenvectors of $\, S_{ij} = \frac{1}{2} (\partial u_i / \partial x_j + \partial u_j / \partial x_i)$ (as in Figure \ref{fig:alignment}a), as they are thought to be the primary cause of non-linearity in vorticity dynamics \citep{Hamlington2008}. 
As such, there has been an effort to link the energy cascade with the geometry of \A\!, particularly as it is implicitly assumed in many sub-grid-scale turbulence models that the velocity-gradient at resolved (large) scales can be used to determine energy transfer at the unresolved (small) scales \citep{Smagorinsky1963,Vela-Martin2021}. We point the reader to the excellent recent review by \citet{Johnson2024} for a more nuanced discussion of the current research, but we will highlight some of the more pertinent works here. 

The eigenvectors of \s define its principal axes, which are also the directions of extremal strain. They are defined as \eone, \etwo, \etri, where \eone and \etri are always the most extensive and compressive vectors, respectively. 
Conversely, for incompressible flows \etwo can either be extensive or compressive depending on the local flow. 
One of the most consistent, but perhaps counterintuitive, findings in this field has been that  \w has a prevalence to align with  \etwo (also shown to be predominately extensive), and not \eone\!, which would result in the maximal amount of vortex-stretching \citep{Betchov1956, Ashurst1987,Tsinober1992,  Hamlington2008,Buaria2020}. 
These same works have also reported the more slight preference for \w to orient itself perpendicular to the most compressive strain axes, \etri. 
An illustration of this preferred alignment is shown in Figure \ref{fig:alignment}a.
The prevalence of these findings across theory, direct numerical simulations (DNS), experiments and a range of turbulent flows lend credence to the idea that these vortex alignment preferences may be considered a universal feature of turbulent flows.

Despite the predominance of \w alignment with \etwo, other works provide arguments as to why it is nevertheless the alignment of \w and \eone that is the main driver of enstrophy generation and energy cascade \citep{Hamlington2008,Buxton2010, Buaria2021}. Looking at the individual contributions of each strain-rate eigenvector to enstrophy-production,  some found the contribution from \eone to be slightly larger than that of \etwo, with a difference that increased with Reynolds number, $Re$  \citep{Jimenez1993,Buaria2019,Zhou2021}. 
\citet{Hamlington2008} and \citet{Buaria2021} showed that the \w-\eone alignment was achieved after removing the local, self-induced, strain-rate, while \citet{Xu2011} showed that this alignment occurred at some characteristic lag time. 
Other authors use the joint probability distribution of the second and third velocity gradient invariants ( a ``QR plot'') to demonstrate the prevalence of vortex stretching over vortex compression. \citet{Buxton2010} combined both approaches and found the \w- \s alignment for each region of QR space. While the alignment behavior of \etwo stayed the same throughout, areas with $R < 0$ displayed \w- \eone alignment. 
Thereby they argued that the extensive \eone alignment is more important for determining enstrophy-production or destruction, and that the universally flat distribution of \eone alignment in other works is due to the summation of the different behaviors throughout QR space. 

On the other hand, many other works propose that the alignment of \w and \s is over-emphasized and instead the self-amplification of strain-rate is most important \citep{Vincent1994,Tsinober1998, Tsinober2009, Carbone2020,Johnson2020,Johnson2021, Vela-Martin2021}.  
For instance \citet{Vincent1994} presented evidence that enstrophy-production is actually driven largely by the shear instability of vortex sheets. 
Similarly \citet{Tsinober2009} showed the cascade is caused more via the compression of fluid elements and strain-self-amplification than by vortex stretching. A key to these arguments is the non-local nature of \w- \s interactions, the direct relationship between strain and energy dissipation, and the highly non-linear nature of strain dominated regions \citep{Tsinober1998, Tsinober2009}. 
Several related works used the the QR space to claim that most nonlinear flow behavior occurs in areas with high strain-production, while nonlinearity is relatively repressed in regions of high vorticity \citep{Tsinober1998, Gulitski2007, Tsinober2009}. 
Further, while strain-production has been found to only depend on strain, vortex stretching has been shown to actually deplete strain-production \citep{Betchov1956, Tsinober2009}. 
Interestingly, \citet{Buaria2019} and \citet{Donzis2008} also found that while extreme strain events were co-located with equally large enstrophy events, intense vorticity events were accompanied by relatively less strain. 
The fact that enstrophy and strain share the same mean value, yet enstrophy appears more intermittent is thought to be due to the fact that vortex stretching amplifies vorticity while depleting strain \citep{Buaria2019, Ishihara2009}. 

The behavior of the velocity-gradient during ``extreme events" in the energy cascade is also of much interest.  The link between \A morphology and extreme energy transfer events can be easily made by considering that in the 1D Burgers equation, strain-self-amplification is responsible for singularity and shock formation in the inviscid limit \citep{Johnson2024}. 
This idea was extended to 3D with the so called ``Restricted Euler" equation, which has been shown to produce finite time singularities in the absence of the anisotropic pressure Hessian and viscosity \citep{Vieillefosse1982, Vieillefosse1984}. 
Notably, the morphologies which ought to bring about this singularity are those found ubiquitously in turbulent flows:  two extensional strain directions (\eone, \etwo) and preferential \w- \etwo alignment, \emph{i.e.} Figure \ref{fig:alignment}a. 
Strain-self-amplification (illustrated in Figure \ref{fig:alignment}b) occurs strictly due to the positivity of \etwo, and it's predicted that the singularity will occur along the region in QR space affiliated with that straining mechanism \citep{Vieillefosse1982,Vieillefosse1984,Johnson2021, Johnson2024}. 
These ideas have been supported by other works investigating extreme events in enstrophy, which have simultaneously found enhanced \w- \etwo alignment, and a quasi-2D structure during the event \citep{Jimenez1993,Buaria2019,Buaria2020, Zhou2021}. 
Finally, \citet{Vela-Martin2021} investigated the morphological and probabilistic differences between downscale and upscale energy transfers. They found these to be mainly driven by  strain-dominated energy fluxes, where downscale transfer was most correlated with strain dominated regions.
Additional investigation of the velocity-gradient structure for upscale energy transfer showed that while vortex stretching and compression exists in both cascade directions (albeit in different proportions), strain dominated regions depends strongly on cascade direction.

In this work we will combine approaches used in previous works to address several unanswered questions. Of main interest is how the morphology of the velocity-gradient behaves for the largest upscale and downscale energy transfer events. We will condition morphological statistics on energy transfer amplitude and make use of both the QR and \w- \s alignment methods to probe the behavior. In the next sections we introduce our experimental facility (Section\ref{sec:exp_facil}) and the main quantities of interest (Section\ref{sec:quants}), and then display non-conditioned morphological statistics (Section\ref{sec:Uncond_results}). We then provide the conditioned statistics in Section\ref{sec:Cond_results} and finish by discussing their implications from different perspectives in Section\ref{sec:discuss}. 

%% file: sections/exp_facility.tex
\section{Experimental Giant von Karman Facility}\label{sec:exp_facil}

\begin{figure}
\centerline{\includegraphics[scale=0.124]{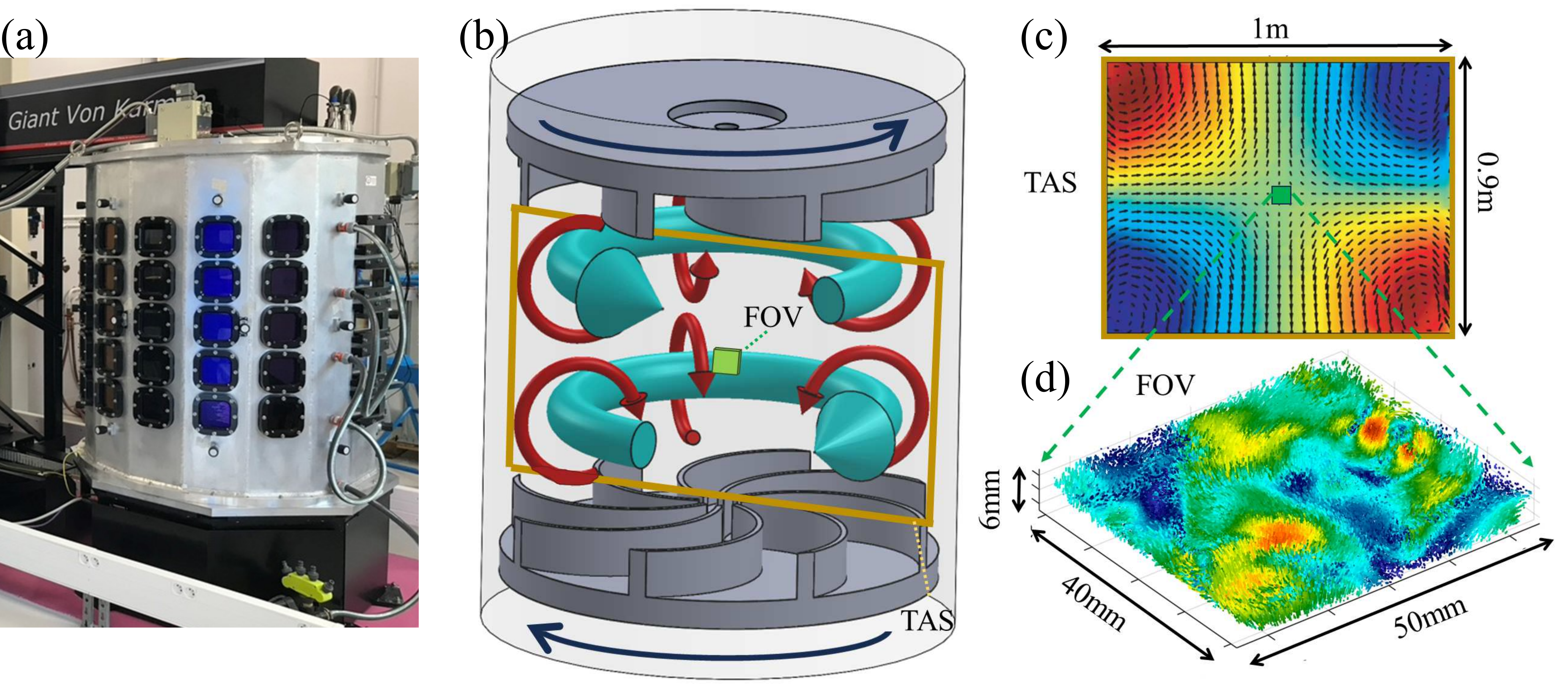}}
\caption{\label{fig:facility} (a) Photo of the GvK experimental facility. (b) Schematic of the facility showing the direction of impeller rotation and time averaged flow. Two toruses revolve in the same direction as their nearest impeller, which induce counter rotating vortices that create the turbulent shear layer where the Field-of-View (FOV) is located. (c) Time averaged slice (TAS) shows the steady-state velocity field in the meridional plane and the relative size and position of the FOV \citep{Cortet2010}. The arrows show the velocity vectors in the axial-radial plane, while color maps to the azimuthal velocity from negative (blue) to positive (red). (d) One instant of the 80 000 particle trajectories found in the FOV. Color indicates velocity magnitude normalized by the impeller tip velocity (0 = blue, 1 = red).  }
\end{figure}

The data presented in this work is purely experimental and is collected from the Giant von K\'arm\'an (GvK) facility at CEA Paris-Saclay. This state of the art facility has been described extensively in previous works, so we resort to a quick summary here \citep{Dubrulle2022,Cheminet2021, Cheminet2022, Debue2021}. 
The facility (Figure \ref{fig:facility}a) consists of a large cylindrical tank with radius ($r$) and height (distance between impellers) of 0.5m and 0.9m, respectively - the radius to height aspect ratio is 1.8. The tank is filled with water that is maintained at a constant temperature of $20^{\circ}C$ by two cooling circuits located above and below the turbines, thus ensuring a statistically constant viscosity ($\nu$) in the flow. 
The flow is driven to a turbulent steady-state via two counter-rotating impellers, which rotate at the same frequency ($F$) and are used to change the integral scale Reynolds number ($Re = 2\pi r^2 F / \nu$). The impellers have 8 curved blades (type TM87 as described in \citet{Ravelet2004}), and are rotated such that the concave portion of the blade advances through the fluid in a ``scooping" direction. 


\begin{table*}\centering 
\renewcommand{\arraystretch}{1.2}
    \begin{tabular}{l c c c c c c c c c c}
    \hline
    \hline
             $\bm{Re}$ &  $F$ [Hz] & $Re_{\lambda}$ & $\langle \epsilon\rangle$ [$\frac{\text{m}^2}{\text{s}^3}$] & $\eta$ [mm] & $\tau_{\eta}$ [ms] &  $T_i$ [s] & $\frac{\tau_{acf}}{\tau_{\eta}}$ & $F_a$ [Hz] &  $\frac{\Delta x_p}{ \eta}$  &  $N$  \\
      \hline  
      $\mathbf{6~269}$   &    0.004  &  87   & 1.89$\times 10^{-7}$ & 1.52  & 2 300 &   39.8  & 3.48  & 50    &  0.3  &   236  \\
      $\mathbf{39~180}$  &    0.025  &  277  & 4.63$\times 10^{-5}$ & 0.38  &  147  &   6.4   & 4.54  & 300   &  0.6  &   416  \\
      $\mathbf{156~719}$ &     0.1   &  616  & 2.96$\times 10^{-3}$ & 0.14  &  18   &   1.6   & 4.63  & 1 200 &  3.0  &   1 856  \\     
      \hline
    \end{tabular}
\caption{Table of relevant experimental parameters for the three $Re$ GvK data sets.}
\label{tab:parameters}
\end{table*}

The resulting flow is fully turbulent with a coherent, large-scale structure in the time-averaged sense (Figure \ref{fig:facility}b-c).
This structure creates a homogenized and quasi-isotropic shear layer in the mid-plane of the cylinder, where the experimental measurements are carried out. The characteristic higher-order turbulent statistics in this area of the flow (\textit{i.e.} Field of View, FOV, in Figure \ref{fig:facility}b-d) have been found to agree well with homogeneous isotropic turbulence (HIT) DNS results \citep{Geneste2019, Debue2018b, Debue2018a,Debue2021}. 

Time resolved 3D velocity fields are obtained using 4D Particle Tracking Velocimetry (PTV), whereby a high-speed laser ($30$ mJ pulse Nd-YLG) illuminates a volume ($V$ = 50 x 40 x 6 mm$^3$) at the centered FOV, which is recorded using four cameras (Phantom Miro m340: 4.1 Megapixel CMOS sensors, with $10~\mu m$ square pixels). The particles used are spherical polystyrene particles  of 5 $\mu$m diameter (Stokes number of $S_{t_{\tau_{\eta}}} \leq 8.1 \times 10^{-5}$ for all cases). 
Particle tracks are obtained using the Davis10 "Shake-The-Box" algorithm \citep{Schanz2016}. 
In house codes are then used to go from Lagrangian particle trajectories to Eulerian velocity fields. This process has been validated and described previously by \citet{Cheminet2021}.
In short, a regularized B-spline filter is first applied on the Lagrangian trajectories to smooth out temporal noise on the tracks. Then a second regularized B-spline interpolation scheme \citep{Gesemann2016} is used to project the velocity field onto a regular Eulerian grid.

The spatial resolution of our data is in turn determined by the number of particles in the FOV ($N_p \approx$ 80,000 on average) where the mean inter-particle distance is: $\Delta x_p = (V / N_p)^{1/3}$. 
When keeping constant flow and diagnostic parameters, increasing/decreasing the rotation frequency has the effect of physically decreasing/increasing the Kolmogorov length ($\eta$) and time ($\tau_\eta$) scales, and thus the effective resolution:  $\Delta x_p / \eta$. 
The simultaneous high $Re$, high resolution nature of this data can be seen in Table \ref{tab:parameters}, made notable by the fact that previous works on the morphology of 3D HIT turbulent data have typically been numerical \citep{Buaria2022}. 
Other relevant experimental parameters displayed in Table \ref{tab:parameters} are defined below. 
\, $\langle \epsilon \rangle$ is an estimate for the average dissipation-rate, found using torque-meters located on the impeller shafts \citep{Kuzzay2015}. 
The integral timescales is $T_i$, while the Taylor Reynolds number ($Re_{\lambda}$) was computed from the measured rms velocity, assuming homogeneity. The relevant frequencies shown are the data acquisition frequency and impeller rotation frequency, $F_a$ and $F$ respectively. 
\rev{The ratio of $\tau_{acf} / \tau_{\eta}$ is the mean de-correlation time of the time-resolved velocity fields relative to the Kolmogorov timescale. }
As this work is statistical and the data acquisition is time resolved, we sample our velocity fields to produce snapshots that are statistically independent temporally. 
The time-step between independent snapshots is determined by $\tau_{acf}$, and the resulting number of statistically independent flow-fields for each case is $N$. 
\rev{The autocorrelation function ($ACF_{ii}(\tau) = \langle v_i(t) v_i(t+\tau) \rangle/ \langle v_i^2(t) \rangle$) is computed at an array of points in the measurement volume for each, mean subtracted, Eulerian velocity component, $v_i$. The de-correlation time $\tau_{acf}$ is then found via $\tau_{acf} = \langle ACF(\tau) = 0.025 \rangle$, or in other words the average de-correlation time of the velocity $ACF$.}


%% file: sections/results.tex
\section{\label{sec:quants} Quantities of Interest}

The analysis of the experimental data revolves primarily around three quantities and their products. These are the strain-rate tensor \s\!, the vorticity vector \w\!,  and the scalar parameter ``$\mathcal{D}_\ell$" (defined below in Equation \ref{eqn:DR}). Again, note that we have sampled the time resolved data to obtain statistically independent data. 

Due to their importance in the production of vortex stretching and dissipation, the alignment of \w and the principal axes of \s are of interest. 
The eigenvectors of \s (\eone, \etwo, \etri) have three corresponding eigenvalues which are defined such that \lone $\, \geq$ \ltwo $\, \geq$ \ltri. 
To find the degree of \w- \s alignment, we consider the alignment cosine, $C_i$, between the vorticity unit-vector, $\hat{\bm{\omega}}$, and the three principal axes of \s such that
\begin{equation}
\label{eqn:cos}
    C_i = | \cos( \boldsymbol{e_i \cdot \hat{\omega}} ) | .
\end{equation}

\noindent $C_i$ is then bounded between 0 and 1 for each principal direction, with \w being perpendicular or parallel to $\bm{e_i}$ for values of 0 and 1, respectively.

Incompressible continuity ($\nabla \cdot \mathbf{u}$ = 0) imposes an additional constraint on the eigenvalues of \s such that \lone + \ltwo + \ltri = 0. Thus, as the most extensive strain eigenvector \eone is always positive (\lone $>$ 0 ) and the most compressive strain eigenvector \etri is always negative (\ltri $<$ 0), the value of \ltwo will change sign depending on the local flow behavior. We quantify the sign and relative amplitude of \ltwo using 
\begin{equation}
\label{eqn:beta}
    \beta = \sqrt{6}\lambda_2 \, / \sqrt{\lambda_1^2 + \lambda_2^2 + \lambda_3^2} . 
\end{equation}

\noindent We will also report statistics on the second and third invariants of the velocity gradient tensor, $Q$ and $R$ respectively. Here $Q= \frac{1}{4} \omega_i \omega_i - \frac{1}{2} S_{ij}S_{ij}$ and $R= -\frac{1}{3} S_{ij}S_{jk}S_{ki} - \frac{1}{4} \omega_i S_{ij} \omega_j$ - the quantities whose joint distributions make up the aforementioned QR plot. Note that velocity derivatives are computed from the Eulerian field using a second-order central-differencing scheme.

Finally, as we are interested in determining the morphology of the flow field around events which transfer large amounts of energy, we measure this energy transfer using the $\mathcal{D}_\ell$ parameter, a term generated in the ``weak Karman-Howarth-Monin equation" \citep{Dubrulle2019}. This is a scale and space dependent energy balance for the point-split kinetic energy, $E^{\rm \ell}(\bm{x}) \equiv u_i u_i^\ell/2$, developed in previous works by \citet{Onsager1949, Duchon2000,Dubrulle2019}.  \, $E^\ell (x)$ changes across time, space, and scale ($\ell$) due to the transport of energy within the flow. In this work we are only interested in one term from the evolution equation of $E^\ell (x)$, the inter-scale energy flux:

\begin{equation}
\label{eqn:DR}
\mathcal{D}_\ell =  \frac{1}{4}\int \nabla \phi^\ell (\xi)\cdot \delta \bm{u} (\delta \bm{u})^2 \; d^3\bm{\xi}\, .
\end{equation}

 \noindent Here the velocity increment, dependent on space ($\bm{x}$) and increment-distance ($\bm{\xi}$), is $\delta \bm{u} = \bm{u}(\bm{x}+\bm{\xi}) - \bm{u}(\bm{x})$. 
 $\phi^\ell$ is a smooth, even, non-negative, and spatially localized smoothing operator which effectively removes fluctuations on scales smaller than $\ell$, meaning smaller $\ell$ results in less smoothing. 
 In our case it is the following Gaussian function: $\phi^\ell(x) = \exp \left( -\frac{30x^2}{2\ell^2} \right) / \left( \frac{2\pi\ell^2}{30} \right)^{1.5}.$  
 The scalar $\mathcal{D}_\ell$ is the non-viscous inter-scale transfer term, \textit{i.e.} the local (in space and time) rate of energy transfer from scales larger than $\ell$ to scales smaller than $\ell$. It adopts the convection that positive $\mathcal{D}_\ell$ values indicate energy transfer to scales below $\ell$ and vice-versa \citep{Dubrulle2019, Debue2021}.
 In the limit of $\ell \to 0$, $\mathcal{D}_\ell$ can also be thought of as an inertial dissipation arising from the irregularity of the velocity field, or in other words, the non-viscous contribution to dissipation due to velocity roughness \citep{Duchon2000}.

 \begin{figure*}
        \centerline{
        \includegraphics[width=\textwidth]{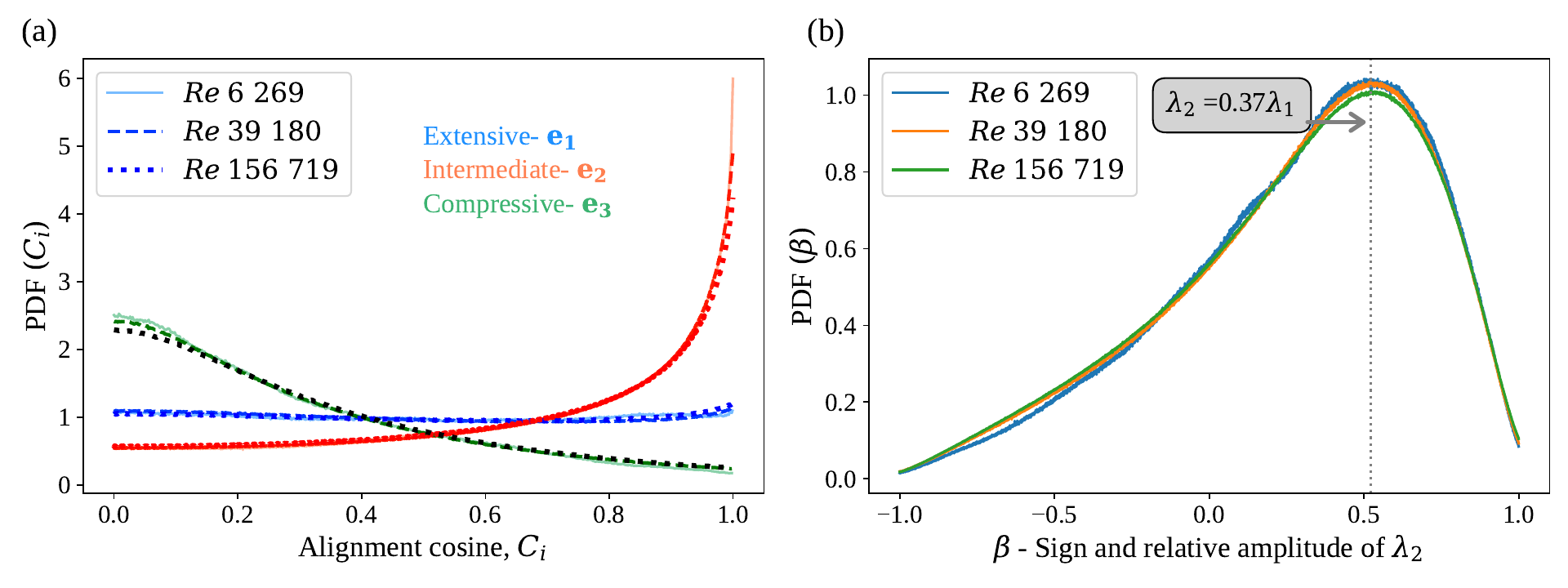}}
    \caption{\label{fig:Uncond}Distributions of $C_i$ and $\beta$ for a range of $Re$. (a) Alignment cosine ($C_i$) distributions of the strain-rate tensor eigenvectors and the vorticity vector. For each eigenvector, different $Re$ are indicated by shades of the same color. (b) Distributions of the relative amplitude of the intermediate \s eigenvector ($\beta$) for different $Re$. The value of \ltwo (relative to \lone) corresponding to the most probable $\beta$ value is shown in the gray box.}
\end{figure*}

\section{\label{sec:results} Results}
\subsection{\label{sec:Uncond_results} Unconditioned Statistics }

Figure \ref{fig:Uncond} shows the unconditioned probability density functions (PDFs) of the alignment cosine ($C_i$ - Equation \ref{eqn:cos}) and the intermediate eigenvector sign and amplitude ($\beta$ - Equation \ref{eqn:beta}) for a range of $Re$. 
The results agree well with the vast amount of previous works, a majority of which were done using DNS on HIT flows. 
We observe the common trend of \w alignment with the intermediate \s eigenvector, which is generally positive (extensive). We also observe the lack of any preferential alignment with the extensive eigenvector (\eone) and a preference for perpendicularity between \w and \etri, the principal compressive axis of \s.  
See Figure \ref{fig:alignment}a for a schematic showing this preferred nominal alignment of an idealized vortex in a strain field.  
While the preference for \ltwo $>0$ was shown by \citet{Betchov1956} to be necessary for net enstrophy-production in a turbulent flow,  the predominant \w- \etwo  alignment has been used to explain the limited vorticity growth rates, as compared to those of passive scalars, observed in turbulence  \citep{Elsinga2010}. 

\begin{figure}
\centerline{\includegraphics[scale=0.6]{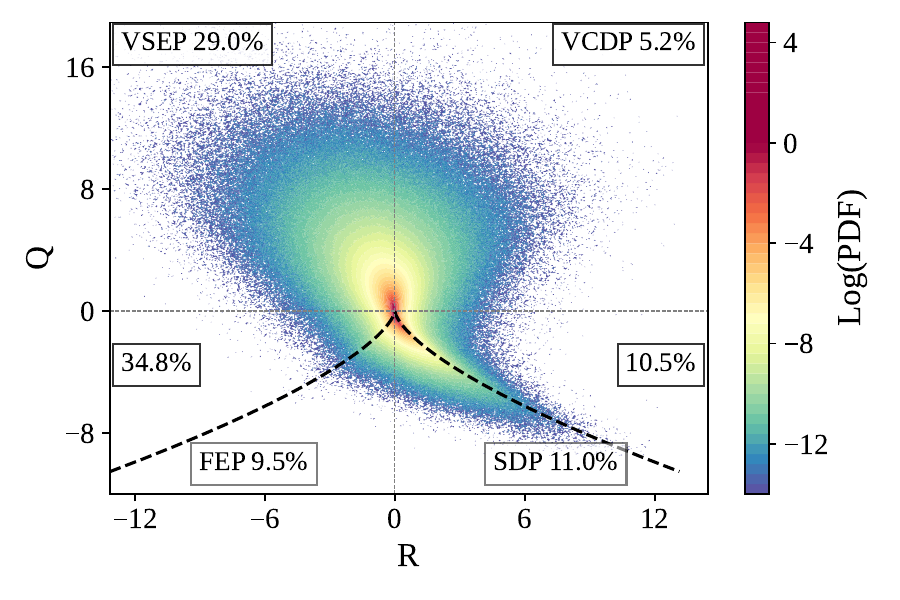}}
\caption{\label{fig:QR} QR plot of the non-conditioned turbulent flow at $Re$=156 719. The dashed line is the Vieillefosse line. VSEP (Vortex-Stretching Enstrophy-Production), VCDP (Vortex-Compressing Dissipation-Production), SDP (Sheet Dissipation-Production) and FDP (Filament Enstrophy-Production) indicate different topological regions of QR space. The percentage of total events in each region are also indicated by the text boxes. \rev{The plots have been non-dimensionalized using the appropriate powers of $\langle S_{ij} S_{ij} \rangle $}.}
\end{figure}

The unconditioned joint PDF of the velocity-gradient invariants $Q$ and $R$ for the case of $Re = 156~719 $, scale $\ell = 6.5\eta$, are shown in the QR plot of Figure \ref{fig:QR}. 
Again due to incompressibility, the first invariant, $P$, of the velocity gradient is zero. 
As such, one can define a line ($D$) which separates the real and complex roots of the velocity-gradient tensor as: $D = Q^3 +(27/4)R^2$. 
Meaning that $D>0$ represents swirling flow topology, while regions with $D<0$ are strain dominated. This is shown as the dashed line in Figure \ref{fig:QR} and is better known as the ``Vieillefosse line/tail''\citep{Vieillefosse1984}.  
The QR plot can be further partitioned based on the sign of $R$ and $Q$, where positive and negative $R$ indicate unstable and stable solutions, respectively, while positive and negative $Q$ indicate enstrophy dominated and strain dominated regions, respectively \citep{Chevillard2006, Chevillard2008,Elsinga2010}. 
One can also interpret the sign of $R$ morphologically, where $R <0 $ indicates a stretching principal direction (i.e vortex-stretching) and $R > 0$ indicates a compressive principal direction (strain-self-amplification or vortex-compression). 

The QR plot can therefore be divided into 4 (or 6) regions, each corresponding to a solution-type of the characteristic equation of the velocity-gradient tensor's eigenvalues, which correspond to different flow topologies. 
The top left region of the QR plot in Figure \ref{fig:QR} labeled VSEP (Vortex-Stretching Enstrophy-Production: $R < 0,\; Q > D$) is characterized by enstrophy producing regions of the flow driven by stable vortex-stretching topologies.
The top right region labeled VCDP (Vortex-Compressing Dissipation-Production: $R > 0,\; Q > D$) contains regions of the flow that produce dissipation via unstable vortex-compression topologies. 
Correspondingly, the bottom right region labeled SDP (Sheet Dissipation-Production: $R > 0,\; Q < D$) pertains to unstable saddle-node, or sheet, topologies which tend to produce dissipation. 
Finally, the bottom left region labeled FEP (Filament Enstrophy-Production: $R < 0,\; Q < D$) produces enstrophy via stable-node, or filament topologies \citep{Debue2021,Chong1990, Chevillard2006, Chevillard2008,Elsinga2010,Danish2018}. 
We direct the reader to the work of \citet{Suman2010} for a more nuanced explanation of stability/instability in the QR plot. 
One can additionally split the VSEP and VCDP regions into $Q > 0$ and $0 > Q > D$ where the former contains swirling topologies in regions dominated by enstrophy, and the latter, while still containing swirling topologies, are regions where strain is more prevalent \citep{Danish2018}.

\begin{figure}
\centerline{\includegraphics[scale=0.5]{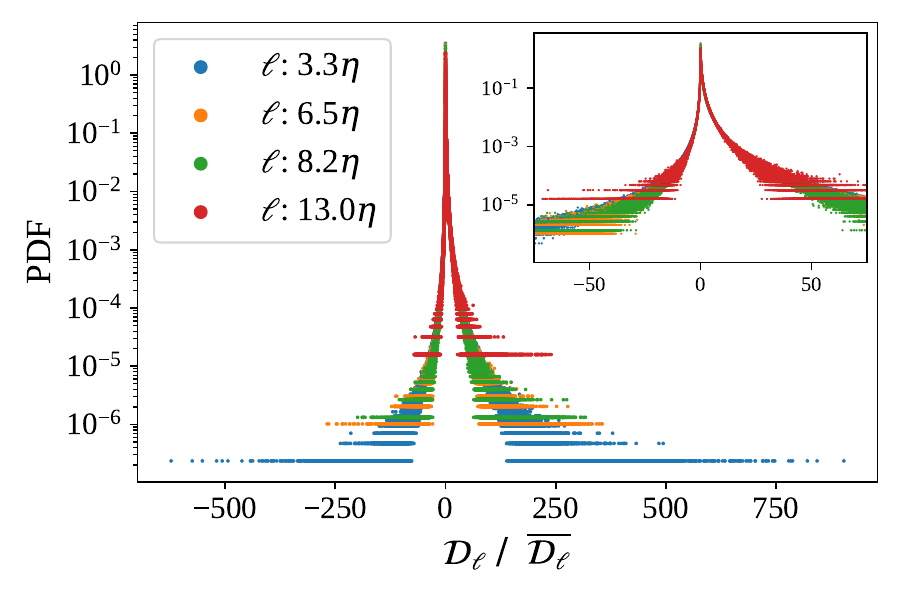}}
\caption{\label{fig:DR} PDFs of mean-normalized energy transfer ($\mathcal{D}_\ell$) to and from scale $\ell$ for a range of scales at $Re$=156 719. \rev{The mean to RMS ratio of $\mathcal{D}_\ell$ at each scale are as follows: 3.3$\eta$ - 0.0032/0.0167,  6.5$\eta$ - 0.0106/0.0421, 8.2$\eta$ - 0.0147/0.0542, 13.0$\eta$ - 0.0255/0.2242. }}
\end{figure}
The resulting tear-drop shape of the QR plot (with the pointed portion along the positive/right Vieillefosse tail) in Figure \ref{fig:QR} is now one of the most notable features in a turbulent flow \citep{Elsinga2010}. 
This familiar shape indicates that, on average, turbulent flows generate enstrophy via vortex-stretching and produce dissipation via sheets and compressive vortices in strain dominated regions.  The ubiquity of the results shown in Figures \ref{fig:Uncond} and \ref{fig:QR} may be interpreted as universal statistical manifestations of the structure of turbulent flows \citep{Tsinober2009}. Additionally, one should note that a random, divergence-free flow field would result in a shape symmetric about the R axis \citep{Elsinga2010}, which is of course not the case here, highlighting the very non-Gaussian nature of turbulence.  

Finally, we quantify the scale-to-scale energy transfer in our flow using $\mathcal{D}_\ell$, previously defined in Equation \ref{eqn:DR}. 
The distribution of $\mathcal{D}_\ell$ is shown in Figure \ref{fig:DR} for a range of scales at the highest $Re$ (156 719). 
One can see the highly non-Gaussian nature of the distribution, as the long tails indicate energy transfer events up to three orders of magnitude higher than the mean. 
Three things should be noted regarding Figure \ref{fig:DR}. The first is that the relative amplitudes of the largest events increase as the scale decreases. 
Second, positive values of $\mathcal{D}_\ell$ (which we will call \drp) indicate ``direct" cascade or downscale energy transfer events, while negative values (\drn) indicate ``inverse" cascade or upscale energy transfer events.  
One should also note the PDF of $\mathcal{D}_\ell$ is skewed towards positive values for each scale. This shows that $\langle \mathcal{D}_\ell \rangle > 0$, which is of course a manifestation of the turbulent cascade - on average, energy flows from larger scales to smaller scales. 
Lastly, we highlight the scale size indicated in terms of $\eta$ in the legend of Figure \ref{fig:DR}.
It is postulated that the transition between the inertial and viscous range occurs in the range of 10-20 $\eta$, placing this work squarely at or below this regime \citep{Danish2018}.

\subsection{\label{sec:Cond_results}Statistics conditioned on extreme energy transfer}

The results presented in Figures \ref{fig:Uncond}-\ref{fig:DR} were produced by sampling the whole flow field. As such, they do not distinguish between intermittent and quiescent regions, or areas of upscale and downscale energy transfer. 
To determine the morphology of the flow during the largest amplitude energy transfers we bin the flow field by decades of positive/negative $\mathcal{D}_\ell,$ and then calculate the local \w- \s  alignment at all points in each respective $\mathcal{D}_\ell$ bin. 
Here we consider extreme/large energy transfer events to be those with amplitudes between 100-1000 times the respective means of \drn and \drp.  
Note that, unless explicitly noted, the conditioning was done for the case of $Re =$156 719 at scale $\ell = 6.5\eta$ as this case provided the most converged statistics. Thus we will drop the $\ell$ from the subscript of $\mathcal{D}_\ell$.

\begin{figure*}
        \centerline{
        \includegraphics[width=\textwidth]{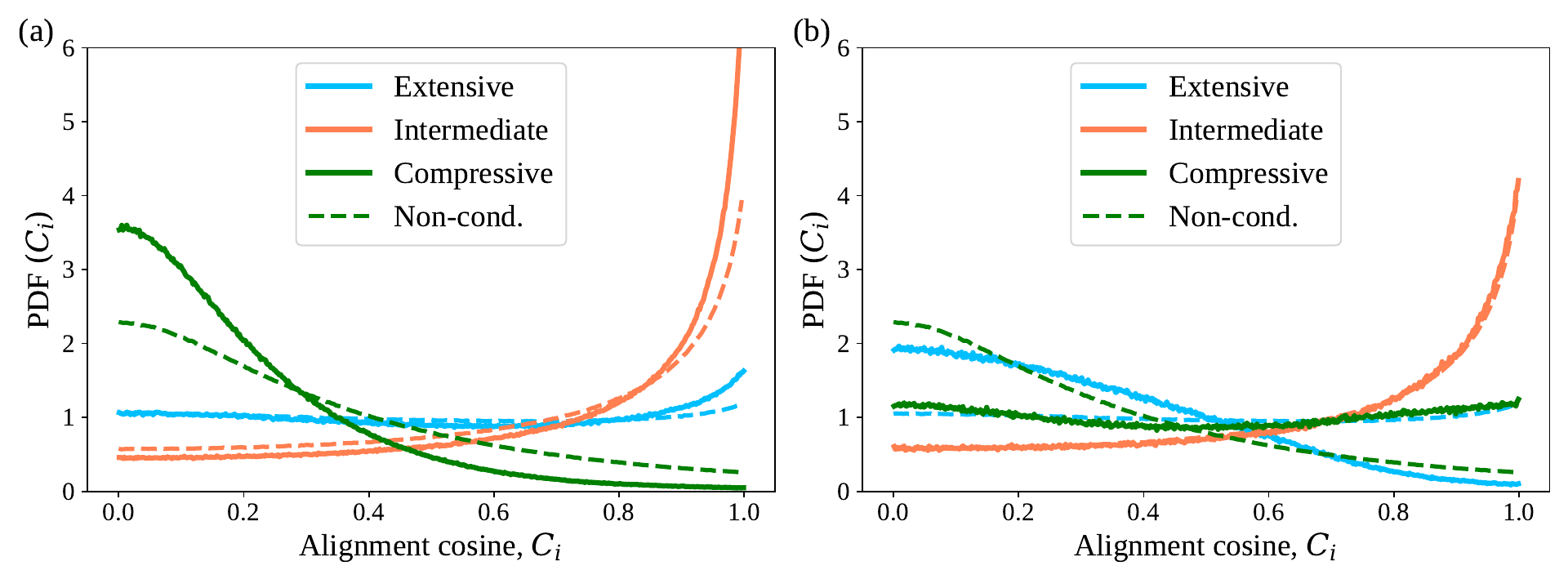}}
    \caption{\label{fig:Cconds}Alignment cosine, $C_i$, conditioned on large magnitude events of downscale and upscale energy transfer. (a) $C_i$ of \w- \s at areas of the flow field with large ( $ > 100$\mdrp) downscale transfer.  (b) $C_i$ of \w- \s at areas of the flow field with large ( $ > 100$\mdrn) upscale transfer. (a-b) Conditioned probabilities for each eigenvector (solid-bold lines) are shown compared to their respective unconditioned alignments (dashed-thin lines). }
\end{figure*}

Figure \ref{fig:Cconds}a shows the alignment cosine $C_i$ conditioned on downscale energy  events satisfying 100-1000\mdrp.
The conditioned probabilities of $C_i$ (solid-bold lines) are shown along with the respective unconditioned alignments (dashed-thin lines). One can see that for large magnitude direct cascade events, the unconditioned \w- \s alignment is enhanced. 
That is to say, \w aligns more strongly with \etwo and has a larger preference for perpendicularity with \etri. Although there is a slight increase in the propensity for \eone to align with \w\!, by and large it remains preferentially unaligned. 
It is particularly notable that the propensity for perpendicular alignment of \w- \etri nearly doubles for the largest direct cascade events. 

\begin{figure*}
        \centerline{
        \includegraphics[width=\textwidth]{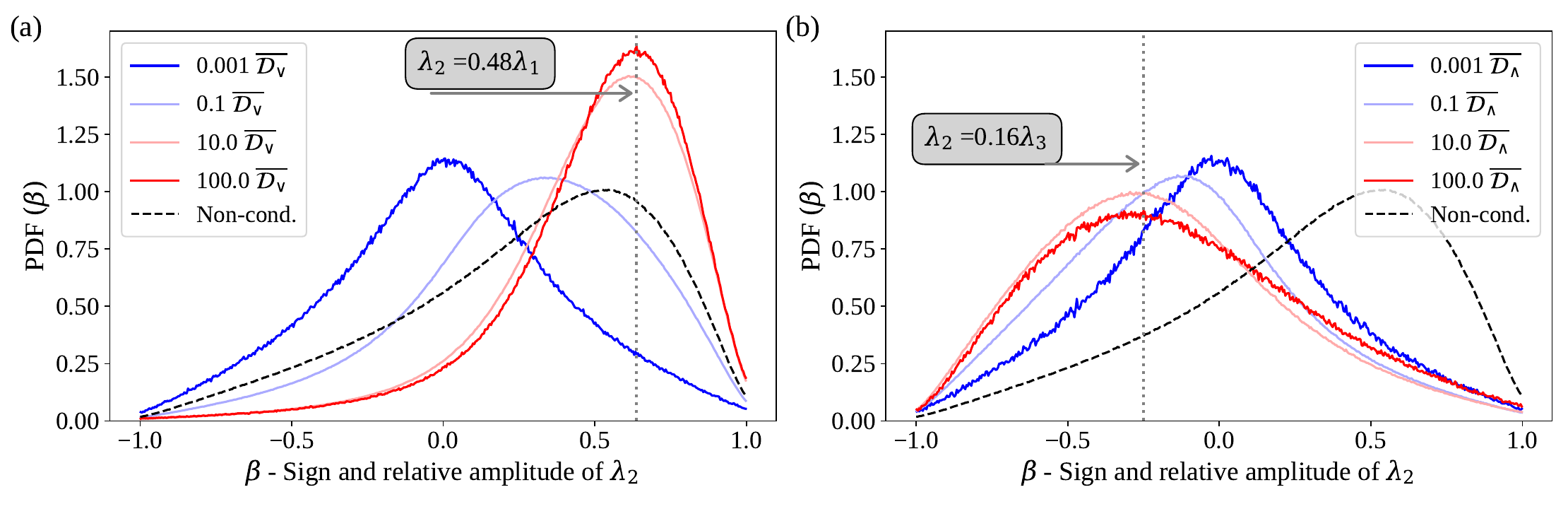}}
    \caption{\label{fig:Bconds} Evolution of the $\beta$ distribution when conditioned on varying levels of downscale (a) and upscale (b) energy transfer amplitude. Conditioning level varies between events of (a) 0.001-100\mdrp \, and (b) 0.001-100\mdrn, from the smallest (bold blue line) to largest (bold red line). The black dashed line is the non-conditioned result. The value of \ltwo (relative to \lone (a) or \ltri(b)) corresponding to the most probable $\beta$ is shown in the gray text box.}
\end{figure*}

Additionally, Figure \ref{fig:Bconds}a shows the distribution of $\beta$ for a range of conditioning levels, from 0.001-100 \mdrp. 
The lowest magnitude downscale energy transfer produces a normal-like distribution in $\beta$ with no preference for compressive or extensional \etwo. 
The distribution of $\beta$ gradually shifts rightward as the level of conditioning on \drp \, increases, such that at the largest downscale transfers ($>100$ \mdrp) \etwo is almost entirely extensional. There is also a notable increase in the most probable magnitude relative to \lone (from 0.37\lone to 0.48\lone), as indicated by the gray text boxes in Figures \ref{fig:Uncond}b and \ref{fig:Bconds}a. 

\begin{figure}
\centerline{\includegraphics[scale=0.55]{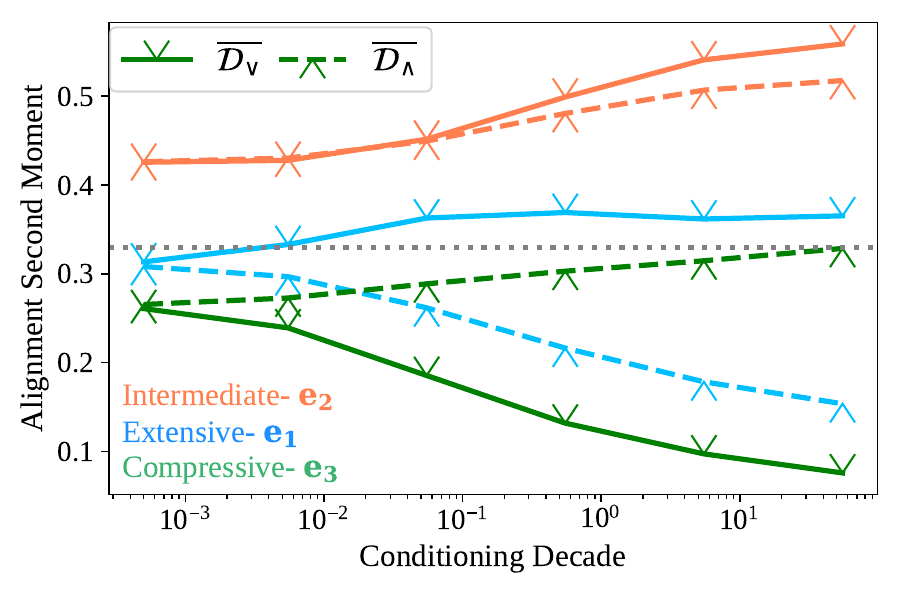}}
\caption{\label{fig:scnd} Average second moment of the alignment cosine across a range of increasing amplitude downscale (\mdrp - solid lines) and upscale (\mdrn - dashed lines) energy transfer conditioning levels. The faint, dotted gray line is at 1/3, which corresponds to uniform distributions of $C_i$. Values closer to 0 or 1 indicate average perpendicular or parallel \w- $\bm{e_i}$ alignment, respectively.   }
\end{figure}

The alignment statistics conditioned on upscale energy transfer, \drn, are shown in Figures \ref{fig:Cconds}b and \ref{fig:Bconds}b. 
The \w- \s alignment  is shown in Figure \ref{fig:Cconds}b, where the \w- \etwo alignment is shown to be effectively identical to the non-conditioned case. However, notably it displays a preference for orthogonality with the extensive  \eone, instead of the compressive \etri as in the non-conditioned case. 

The evolution of the $\beta$ distribution is again shown, but for increasing upscale transfers (0.001-100 \mdrn) in Figure \ref{fig:Bconds}b. Once again, we note the effectively normal distribution of $\beta$ values for the smallest \drn $\,$events. 
However, as the conditioning level increases, the distribution of $\beta$ moves leftwards to negative values.
At the largest upscale energy transfers, $\beta$ is primarily negative (compressive), albeit with a wider distribution that includes a sizable amount of positive \ltwo.  
The relative magnitude of \ltwo\!'s most probable value as compared to its complimentary eigenvalue (\ltri in this case) has also decreased compared to the non-conditioned case - from 0.37\lone to 0.16\ltri. 

A notable feature that applies to both the direct and inverse cascade cases, is that \w always shows a preferentially perpendicular alignment with the largest magnitude eigenvector, or the direction of highest amplitude strain.
For \drp, as \ltwo $>0$, \w is most perpendicular to \etri (because \lone + \ltwo = \ltri ). While for \drn, \, \w is preferentially orthogonal to \eone in the inverse cascade as \lone = \ltwo + \ltri.

In order to gauge how the alignment might change across different energy amplitudes we investigate the average second moment of the alignment cosine, $\langle (\bm{e}_i \cdot \hat{\bm{\omega}})^2 \rangle$. 
This variable is useful in that it is bounded between 0 and 1, and the sum of the three contributing directions will also sum to 1. 
The value for each eigenvector gives an idea of the average alignment for the distribution of $C_i$ at each conditioning level.
Its average value conditioned on varying levels of up/downscale energy transfer is shown in Figure \ref{fig:scnd}. 
One can see that at the highest conditioning levels we recover the results of Figure \ref{fig:Cconds}, where there is greatest preference for \w- \etwo alignment and a orthogonal preference which swaps between \etri for \mdrp \, events (downscale - solid lines) and \eone for \mdrn \, events (upscale - dashed line).  
If we note that a eigenvector with no alignment preference (uniformly distributed) will have a value $\langle (\bm{e}_i \cdot \hat{\bm{\omega}})^2 \rangle$ = 1/3 we can make a surprising observation. 
This is the fact that at the lowest energy transfer levels, the average \w- \s alignment seems to maintain a morphological structure.
This is surprising as several works have found that $\langle (\bm{e}_i \cdot \hat{\bm{\omega}})^2 \rangle$ conditioned on lowest levels of enstrophy or strain will converge to 1/3 for all eigenvectors \citep{Buaria2020}. 
This perhaps points to the fact that energy transfer  is more inherently linked to morphological structure.

\begin{figure}
        \centerline{
        \includegraphics[width=\textwidth]{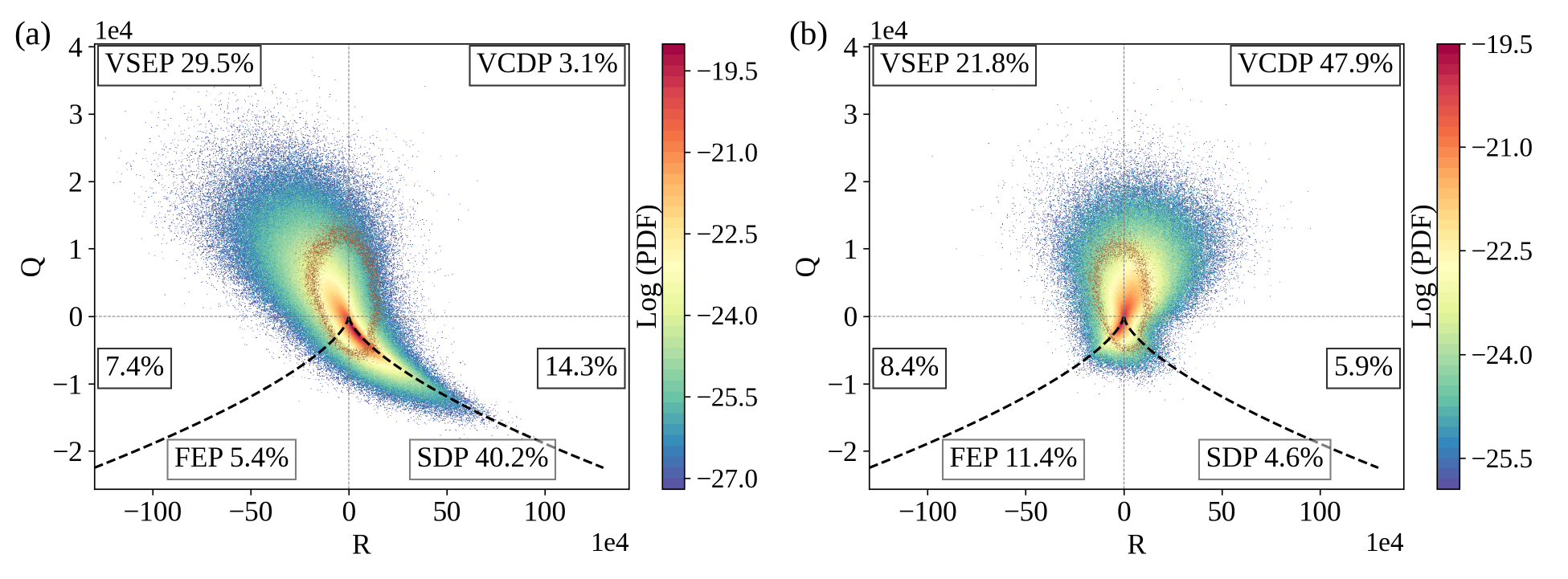}}
    \caption{\label{fig:QRcond}QR plots conditioned on large magnitude (a) downscale ( $> 100$\mdrp) and (b) upscale ( $> 100$\mdrn) energy transfer events. The faint-red ring of scatter-points in the center of each figure shows the outer contour-level for the QR plot conditioned on the mean (a) direct and (b) inverse cascade. The percentage of events located in each labeled region (acronyms defined previously in the text and Figure \ref{fig:QR}) are indicated in the text boxes. The dashed line is the Vieillefosse line. \rev{The plots have been non-dimensionalized using the appropriate powers of $\langle S_{ij} S_{ij} \rangle $}.}
\end{figure}

We next use the conditioned statistics to generate QR plots for the largest upscale and downscale energy transfer. 
Figure \ref{fig:QRcond}a shows the QR plots for downscale energy transfers $>100$\mdrp. While the resulting tear drop is similar to the non-conditioned case, it is considerably more elongated along the Vieillefosse line. 
Thus, a vast majority of its events are in the VSEP region, where vortex-stretching is prominent, and in the sheet forming dissipation producing regions, SDP. The large downscale energy transfer events also pull the pointed portion of the tear drop further down the right Vieillefosse tail, particularly for the most probable events. In regards to the non-conditioned QR plots the greatest changes are in the SDP region (11$\%$ to 40$\%$) and the enstrophy-producing non-swirling region ($35\%$ to 7$\%$).

Alternatively, Figure \ref{fig:QRcond}b shows the QR plot for upscale energy transfers $>100$\mdrn. The tear drop for the inverse cascade is shown beginning to tilt along the opposite Vieillefosse left tail. 
In this case there are proportionately many more events in the VCDP region, corresponding to vortex-compression topology. 
Additionally, while there is a slight hitch remaining along the right Vieillefosse tail, the most probable events are oriented along the left tail. In regards to the non-conditioned QR plot the greatest changes occur in the VCDP region (5$\%$ to 48$\%$), the two regions around the right Vieillefosse tail (totals of 20$\%$ to 11$\%$) and again in the enstrophy-producing non-swirling region above the left Vieillefosse tail ($35\%$ to 8$\%$). 
It should be noted that because net energy transfer is downward, there are fewer statistics for both the inverse cascade as a whole and for large amplitude upscale events, as compared to their downscale counterparts.  Thus we have found that the  \mdrp \, is typically 2.5 times larger than the amplitude of \mdrn. 

Comparatively, in the downscale case most events are in the regions of enstrophy-production via vorticity (VSEP) and dissipation-production via strain (SDP). However in the upscale case, while there is still vortex-stretching producing enstrophy, there is also a notable amount of enstrophy-production via straining and filaments (FEP). 
Interestingly, dissipation-production is now primarily due to vortex-compression (VCDP).
Furthermore, as noted previously there exists a much higher degree of nonlinearity in the direct cascade as opposed to the inverse cascade \citep{Tsinober1998, Tsinober2009}.
As these nonlinearities are highly correlated with strain dominated  ($2S_{ik}S_{ik} > \omega^2$), and strain producing regions ($-S_{ik}S_{km}S_{mi} > 3/4 \omega_i \omega_k S_{ik}$) the largest downscale energy transfer may be highly nonlinear due to the preponderance of events along the right tail of Figure \ref{fig:QRcond}a. 






%% file: sections/discussion.tex
\section{Discussion} \label{sec:discuss}


\subsection{Contribution of Individual Eigenvectors}\label{sec:indiv_cont}
Understanding the amplitude and direction of each eigenvector is crucial for delineating the contributions of vortex-stretching and strain-rate to large amplitude energy transfer. 
These contributions can be better understood by examining the governing equations for the enstrophy, $\Omega$, and an estimate of the dissipation rate, $\Sigma$, (where $\Sigma = 2 S_{ij}S_{ij}$ and the local dissipation rate is $\epsilon = \nu \Sigma$), both shown here:

\begin{equation}
\label{eqn:enst}
    \frac{1}{2} \frac{D\Omega}{Dt} = \underbrace{\omega_i^2 \lambda_i \cos^2{(\omega,\bm{e_i})}}_{\text{VS}} + \nu \omega_i \nabla^2 \omega_i ,
\end{equation}

\begin{equation}
\label{eqn:dissip}
    \begin{split}
        \frac{1}{2}\frac{D \Sigma}{Dt} = -\underbrace{(\lambda_1^3 +\lambda_2^3 +\lambda_3^3)}_{\text{SSA}} - \frac{1}{4} \underbrace{\omega_i^2 \lambda_i \cos^2{(\omega,\bm{e_i})}}_{\text{VS}} - S_{ij} \frac{\partial^2 p} {\partial x_i \partial x_j} + \nu S_{ij} \nabla^2 S_{ij}  .
    \end{split}
\end{equation}

\noindent Above we have neglected external forcing and replaced the vortex-stretching (VS = $\omega_i \omega_j S_{ij}$) and strain-self-amplification terms (SSA = $S_{ij}S_{jk}S_{ki}$) with equivalent terms based on the eigenvalues of \s \citep{Tsinober2009,Buaria2021}.

From Equation \ref{eqn:enst} we can see that only the first term on the RHS (the vortex-stretching term) is a source of $\Omega$ production, and this is only the case when $\lambda_i > 0$.  Additionally, the contribution from each eigenvector depends on both the alignment and the magnitude. 
The VS term also occurs in Equation \ref{eqn:dissip} as a sink, albeit with three quarters less effect. 
Thus, in this case when $\lambda_i >0$, the vortex-stretching term opposes dissipation-production, while it creates dissipation when $\lambda_i <0$. 

\begin{figure}
\centerline{\includegraphics[scale=0.55]{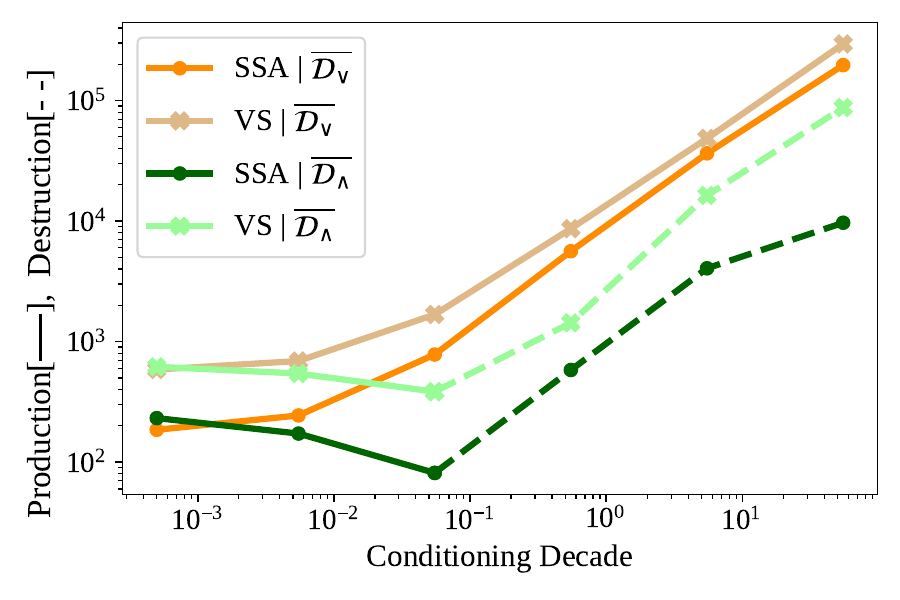}}
\caption{\label{fig:Vs_Ssa} Average vortex-stretching (VS in Equation \ref{eqn:enst}) and strain-self-amplification (SSA in Equation \ref{eqn:dissip}) terms conditioned on increasing amplitudes of direct (\mdrp \,- orange hues) and inverse (\mdrn \,- green hues) cascade events. Dashed portions of the lines represent destruction of the respective quantity. Pale hues show VS, while deeper colors are SSA.}
\end{figure}

Alternatively, the first term on the RHS of Equation \ref{eqn:dissip} is the strain-self-amplification term, which produces dissipation only if two eigenvalues are positive. 
This has two important implications, the first being that due to incompressibility \lone + \ltwo + \ltri = 0, and in the situation where \lone, \ltwo $>0$, the only term contributing to the production of dissipation through SSA is the compressive strain eigenvector \etri \citep{Gulitski2007}.  
One should also note that in this case (\ltwo$>0$), the larger the magnitude of \ltwo, the larger the SSA, as \lone$\approx$\ltwo minimizes their respective negative contributions, due to the cubed powers in Equation \ref{eqn:dissip}.  
Second, it is not possible to produce dissipation through the SSA term when \ltwo$<0$ as only two positive eigenvalues yield a positive number, and by definition \lone$>0$ and \ltri$<0$. 
This can be observed in our data in Figure \ref{fig:Vs_Ssa}, as the SSA term conditioned on the inverse cascade (SSA$|$\mdrn) begins to destroy dissipation once the upscale energy amplitude reaches a level where the distribution of \ltwo skews negative (see $\beta$ in Figure \ref{fig:Bconds}b).
This fact demonstrates that the compressive action of \etri alone is what skews $\langle S_{ij}S_{jk}S_{ki}\rangle$ negative, and because this term is proportional to the distribution of velocity derivatives, it's also responsible for their negative skewness. 
In other words, the compression of \etri  results in the negative skewness of longitudinal velocity increments, a  hallmark of the intrinsic non-Gaussian structure of turbulent statistics, and an exact result of Kolmogorov 4/5ths law \citep{Rosales2006}. 
This suggests that because we observe a generally negative \ltwo (Figure \ref{fig:Bconds}b) in the inverse cascade case, longitudinal velocity increments should skew less negatively for large upscale energy transfer. 

The average values of the VS and SSA terms in Equations \ref{eqn:enst} and \ref{eqn:dissip} are shown for different levels of direct and inverse cascade conditioning in Figure \ref{fig:Vs_Ssa}. 
For the direct cascade case, both VS and SSA increase as a power-law dependent on the energy transfer amplitude. 
However, in the case of inverse cascade, these terms both decrease as upscale amplitude increases, until the point where they actually become net sinks of enstrophy and dissipation (as indicated by the dotted lines in Figure \ref{fig:Vs_Ssa}).

\begin{figure}
\centerline{\includegraphics[scale=0.55]{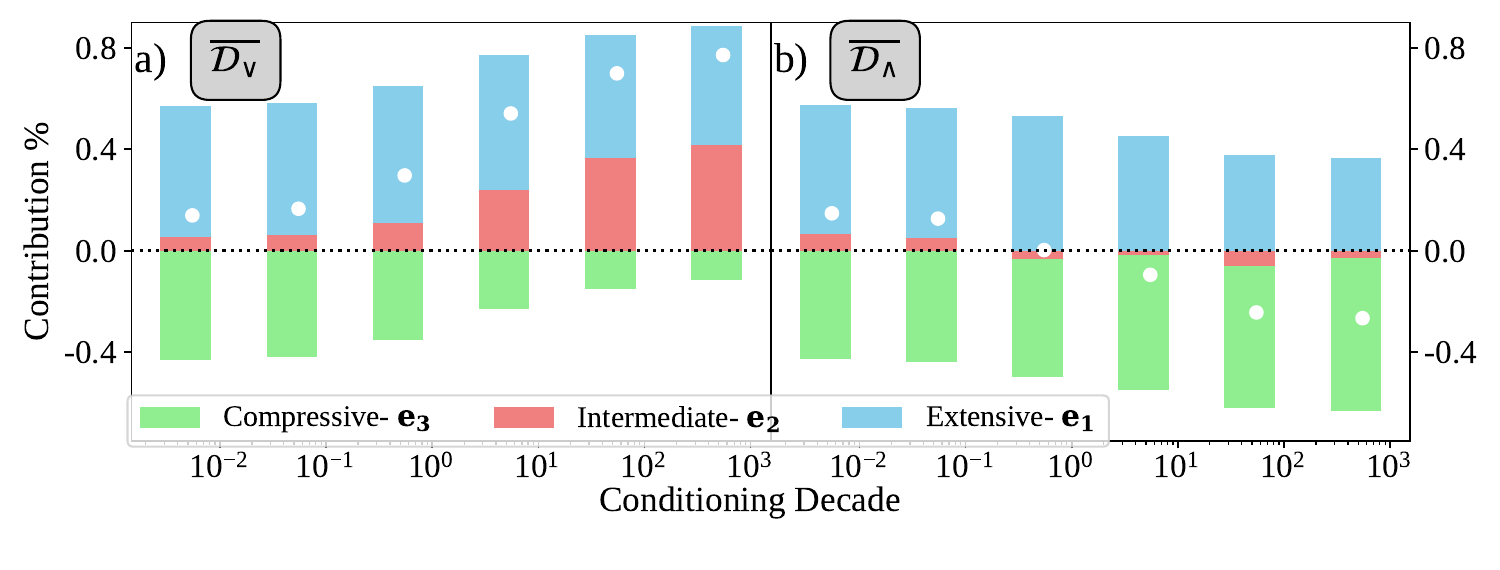}}
\caption{\label{fig:EnstTerms} Normalized contributions of each \s eigenvector towards the vortex-stretching (VS) term in Equation \ref{eqn:enst} conditioned on increasing amplitudes of (a) \mdrp - downscale and (b) \mdrn - upscale energy transfer. \rev{The bars shown are all the same normalized height, and the size/contribution of each eigenvector is found by dividing the absolute value of the eigenvalue contribution by the absolute value sum of all the contributions. The eigenvalue contributions which contribute negatively are shown below the x-axis. The white dot signifies the net normalized sum of the contributions}  }
\end{figure}

Understanding the contributions to the VS term are more complicated due to the simultaneous dependence on alignment, sign, and amplitude. 
Thus we show the average individual contributions of each eigenvector to enstrophy-production via the normalized vortex-stretching term in Figure \ref{fig:EnstTerms}.
One can see that in the direct cascade case of Figure \ref{fig:EnstTerms}a, the \etwo contribution grows steadily with increasing energy transfer amplitude until the largest events where it nearly matches the contribution from \eone. 
Notably, \eone is shown to still contribute majorly despite the preference of \w- \etwo alignment, at least in the average sense. 
Concerning the inverse cascade (Figure \ref{fig:EnstTerms}b), we see that once the mean conditioning level is reached, there is actually a net destruction of enstrophy due to VS (as the negative contributions crosses the white $50\%$ line). 
Further, the relative contribution of \etwo remains small for all conditioning levels and the vortex-stretching term in the inverse cascade is dominated by the most extensive and compressive eigenvectors, despite the preference for \w- \etwo alignment. 

\subsection{Importance of Strain vs. Vortex Stretching}
If we return to the conditioned $\beta$ statistics (Figure \ref{fig:Bconds}), during the largest direct cascade events the relative magnitude of \ltwo increases from 37$\%$ of \lone in the non-conditioned statistics to nearly 50$\%$.
Coupled with the stronger degree of \w- \etwo  alignment, this indicates a very large contribution from \etwo to the velocity-gradient dynamics, as we have already remarked in Figure \ref{fig:EnstTerms}a.  
One can also see the near total absence of  negative(compressive) \ltwo for the largest direct cascade events (Figure \ref{fig:Bconds}a), which can also be observed in the corresponding QR plot of Figure \ref{fig:QRcond}a. 
This manifests such that there are merely 3$\%$ of all events located in the vortex-compressing region VCDP.  
Additionally, simultaneous \w alignment with \etwo and orthogonality with a large magnitude compressive \etri indicates morphology that is more akin to strain-self-amplification (\textit{i.e.} Figure \ref{fig:alignment}b) than vortex-stretching. 
This is indeed observed in the QR plot as over 50$\%$ of events are found along the right Vieillefosse tail. 

In Figure \ref{fig:Bconds}b, for the largest inverse cascade events, $\beta$ has the reverse trend, where the relative amplitude of \ltwo decreases from 37$\%$ of \lone to 16$\%$ of \ltri. 
Meaning that for the largest upscale events the compressive and extensive eigenvectors have very big relative amplitudes and dominate the strain magnitude, despite the domination of \w- \etwo alignment.   
One can see the link between the inverse cascade QR plot of Figure \ref{fig:QRcond}b and Figure \ref{fig:Bconds}b by noting the relatively wide distribution of $\beta$ during the largest events. 
Because there is a large number of positive and negative values of \ltwo, this manifests in the QR plot as a plethora of events in both the vortex-stretching (VSEP) and compressing (VCDP) regions.  

Therefore we posit that because there is a comparable number of vortex-stretching topologies in the VSEP region for both the direct (30$\%$ in Figure \ref{fig:QRcond}a) and inverse  (22$\%$ in Figure \ref{fig:QRcond}b) cascade, vortex-stretching is not the primary factor in determining cascade direction. 
Instead it is the number of vortex-compressing events (only 3$\%$ in downscale vs 50$\%$ in upscale) and strain-self-amplification events (over 50$\%$ in downscale and under 10$\%$ in upscale) that mark the key phenomenological differences in cascade direction. 
Additionally,  one can see the most probable values of the large amplitude inverse cascade beginning to orient along the left Vieillefosse tail in Figure \ref{fig:QRcond}b.  

More generally, these results seem to add credence to the idea that the role of vorticity in the energy cascade is overblown \citep{Tsinober1998,  Carbone2020, Johnson2020,Vela-Martin2021}. 
The most obvious sign of this is seen in Figure \ref{fig:QRcond}, as was just discussed above.
The eigenvector directionality for the inverse (\ltri, \ltwo $<0$) and direct (\lone, \ltwo $>0$) cascades found in this work also agree with that found in \citet{Vela-Martin2021}.  This result shows that downscale energy transfer is much more likely to have self-strain-amplification due to its single negative eigenvector. 
While the largest upscale energy transfer events show a repulsion for straining topologies - only around $15\%$ of all events exist below the Vieillefosse lines in Figure \ref{fig:QRcond}b.  
However, a notable departure between this work and that of \citet{Vela-Martin2021} is our observation that the number of vortex-compressing events varies drastically between the large amplitude direct and inverse cascade. 
The aforementioned numerical work (large eddy simulation) showed vortex compression to exist in large quantities for both directions.  We posit that this difference is primarily due to two factors, the first being that our statistics concern scales (6.5$\eta$) much smaller than the large scales explicitly resolved in their work. Second, our statistics are conditioned on the largest energy transfer events, wheres their work artificially forced the average flow of energy to be up or downscale.


\subsection{Morphology and Singularity Aspects}
Recalling aspects of the discussion above can gives us an idea on the vortex morphology during large amplitude energy transfer events. 
Considering that during the largest downscale events, we observe 1) high positivity of \ltwo, 2)  relatively large magnitude \ltwo as compared to \lone, and 3) a strong preference for orthogonality between \w- \etri we see a mechanism for the formation of sheet like structures. 
The vortex, being well aligned with  a large amplitude \etwo, is compressed strongly by the largest amplitude eigenvector \etri, and simultaneously pulled strongly by \etwo and \eone.
This straining mechanism, illustrated in Figure \ref{fig:alignment}b, is associated with the squeezing of fluid elements into sheet-like structures and has been observed in previous works in flow-regions with high vorticity \citep{Vincent1994, Rosales2006}.
The formation of these sheets can then induce shearing instabilities that in turn cause roll up of the compressed vortex sheets. 
The fact that the most populous region of the QR plot in this case is along the Vieillefosse right tail perhaps signals that quasi-singularities may be associated with vortex-sheet morphology more so than those of vortex tubes.

We must also consider that this dominant alignment in the direct cascade case was shown by \citet{Vieillefosse1982, Cantwell1992} to cause Restricted-Euler singularities in the velocity-gradient along the right Vieillefosse tail. 
Notably, we not only see this form accentuated for the largest downscale energy transfer events, but also an elongation of the tear-drop shape along the right Vieillefosse tail in Figure \ref{fig:QRcond}a.
Similar elongation have been previously found for increases in $Re$ \citep{Chevillard2006}.
For the largest amplitude direct cascade transfers, a majority of events are clustered in this region where singularities are expected to occur. 
This may lend credence to the idea that quasi-singularities, or large non-linearities events in the velocity field (shown to occur in greater proportions in this area \citep{Tsinober1998, Tsinober2009})  might be associated with the largest downscale energy transfer events. 
Our results also support the prediction of \citet{Vieillefosse1982} that this region along the right tail should have large negative strain skewness. 
One can see this in two ways, first by observing the power-law increase of SSA for increasing amplitude of direct cascade in Figure \ref{fig:Vs_Ssa}. 
Second, by recalling the increase of \ltwo relative to \lone for the largest downscale energy transfer events in Figure \ref{fig:Bconds}a.
Which, because of incompressibility (\ltri = \lone + \ltwo $\approx$ 1.5\lone $\approx$ 3\ltwo) and the cubed powers in the SSA term of Equation \ref{eqn:dissip} (as already explained in Section\ref{sec:indiv_cont}), causes the increasing dominance of the negative contribution to strain coming from \ltri.
This further associates strain skewness with energy transfer.  
Additionally, the increasing number of events along the Vieillefosse tail for increasingly large downscale transfer events, perhaps signals that the energy cascade is piloted by Euler restricted (singular) dynamics. 

However, while an increase of events along the Vieillefosse tail points to an increase in the velocity-field non-linearities, the dominant morphology may simultaneously point towards the opposite conclusion. 
For the largest downscale events we noted the enhanced parallel and perpendicular alignment between \w- \etwo and \w- \etri respectively. 
In other words, this alignment gives evidence for quasi-2D structures where \w is increasingly parallel to the intermediate strain direction while the other two eigenvectors lie in the equatorial plane.  
This behavior has also been observed previously, but for extreme events in enstrophy, not energy cascade \citep{Jimenez1992, Buaria2020}. This should in turn lead to a reduction in the non-linearity, as the terms \, $\omega_i \omega_j S_{ij}$ \, and \, $\omega_i S_{ij}$ \,both vanish for 2D flows \citep{Tsinober2009}. This would indicate a higher degree of non-linearity during inverse cascade events than in direct cascade events, a conclusion in contrast to that drawn from the QR plots. However, one could argue that although an increased two dimensionality may inhibit non-linearities induced by the interaction of velocity and vorticity, non-linearities existing due to the so called ``turbulent pressure" can still exist and even increase.  
Clearly further work is needed to more explicitly quantify and investigate ``non-linearities" in the velocity field.  

We close by discussing perhaps the most interesting finding of this work. 
If we return our attention to Figure \ref{fig:QRcond}, one should focus on the red portion of the contour plots indicating the location of the most probable events. 
It is notable that these events hug the left (Figure \ref{fig:QRcond}b) and right (Figure \ref{fig:QRcond}a)  Vieillefosse tails for the upscale and downscale cases, respectively. 
In other words, most large magnitude upscale and downscale energy transfer events are approximately symmetric about the R-axis; a signature of symmetry under time reversal ($ t \rightarrow -t, \; \bm{u}\rightarrow-\bm{u}$ ). 
This is noteworthy because the governing Navier-Stokes equations (NSE) should only be invariant in the inviscid limit \citep{Dubrulle2019}. 
While the same R-axis symmetry has been observed in the aforementioned work of \citet{Vela-Martin2021}, it was at large scales, where viscosity can be more readily neglected. 
However in the current work, we are presenting results just above Kolmogorov's dissipative scale (6.5$\eta$); clearly close enough that viscosity ought to break symmetry in the NSE. 
All the same, around events which transfer large amounts of energy up and down scale, we observe evidence of time reversal symmetry. 
It is hypothesized that this is because the flow is inertially dominated in the space surrounding these events, allowing for the assumption of inviscid-ness even at such small scales.

\subsection{Future Work}
Further work is needed to investigate whether there exists a type of universal scaling for extreme events in energy transfer, enstrophy, and strain, such as in \citet{Buaria2022}. 
An investigation such as this would be useful in determining the validity of the multi-fractal model of turbulence.  
It is also of interest to better determine how the cascade is linked to intermittency, and how morphologies of velocity-gradients might change based on their location in or outside of intermittent regions.
We have the ability to probe this  as previous works have used the  $\mathcal{D}_\ell$ parameter to spatially identify regions of high intermittency \citep{Dubrulle2019}. 
Further, a study of the turbulent statistics for only upscale energy transfer regions is in order, as there are several properties worth investigating such as the change in skewness of the longitudinal velocity increments. 
We must also address the issue of scale. While the conditioned statistics presented here were for one specific scale ($\ell =6.5\eta$), it is not know how morphology may change as we approach or exceed the Kolmogorov scale. 
Additionally, for an extreme energy transfer at one scale, it important to know whether there are there imprints of this event felt at scales above/below it.  
Finally, it is of great interest to know how the pertinent morphologies found in this work compare to those generated during vortex-reconnection, as this is another mechanism possibly linked to singularities and extreme energy transfer \citep{Yao2020,Yao2021}.